\newcommand{\PreprintNumber}{Fermilab CONF-99/291-T}

\newcommand{\simle}{\mathrel{\lower0.5ex\hbox{$\buildrel<\over\sim$}}}

\newcommand{\BulletItem}{\newline$\bullet$\hspace{0.4em}\nolinebreak}

\newcommand{\BtoD}{\ensuremath{B \to D\ell\nu}}
\newcommand{\BtoDstar}{\ensuremath{B \to D^*\ell\nu}}
\newcommand{\BtoPi}{\ensuremath{B \to \pi\ell\nu}}

\newcommand{\hAone}{\ensuremath{h_{A_1}(1)}}

\newcommand{\RhoBar}{\overline{\rho}}

\newcommand{\braOket}[3]{%
\ensuremath{\left\langle{#1}|{#2}|{#3}\right\rangle}}

\newcommand{\Vev}[1]{\ensuremath{\left\langle{#1}\right\rangle}}

\newcommand{\Over}[2]{%
\raisebox{0.5ex}{\ensuremath{#1}}/\raisebox{-0.5ex}{\ensuremath{#2}}
}
\newcommand{\etal}{\textit{et al.}}

\newcommand{\JournalRef}[4]{{#1}, \textbf{#2} ({#4}) {#3}}

\newcommand{\XXX}[2]{\href{http://xxx.lanl.gov/abs/#1/#2}{#1/#2}}

\newcommand{\PR}{Phys.\ Rev.}
\newcommand{\PRD}{Phys.\ Rev.\ D}

\newcommand{\NucPhys}{Nucl.\ Phys.}

\newcommand{\PLB}{Phys.\ Lett.\ B}
\newcommand{\ZPhysC}{Z.\ Phys.\ C}
\newcommand{\SoJNucPhys}{Soviet J.\ Nucl.\ Phys.}
\newcommand{\PhysRpt}{Phys.\ Rept.}

\newcommand{\EqnRef}[1]{Eq.~\ref{eqn:#1}}

\newcommand{\FullStop}{.}
\newcommand{\Comma}{,}

\newcommand{\TabRatioCoeffs}{%
\begin{table}
\caption{Ratios and their $1/m_Q^2$ coefficients.}
\label{tab:MsqCoeffs}
\begin{tabular}{lc} \hline
\qquad Ratio       &coefficient $c^{(2)}_i$ \\ \hline \\[-1.6ex]
$\RhoBar_{V_0}\sqrt{R^{B\phantom{^*}\to D\phantom{^*}}_{V_0}}$
      &${\ell_P}/4$ \\
$\RhoBar_{V_0}\sqrt{R^{B^*\to D^*}_{V_0}}$
      &${\ell_V}/4$ \\
$\RhoBar_{A_j}\sqrt{R^{B\phantom{^*}\to D^*}_{A_j}}$
      &$({\ell_P}+{\ell_V}+{\Delta})/8$  \\ \hline
\end{tabular}
\end{table}}

\newcommand{\FigMassDep}{%
\begin{figure}[t]
\epsfig{file=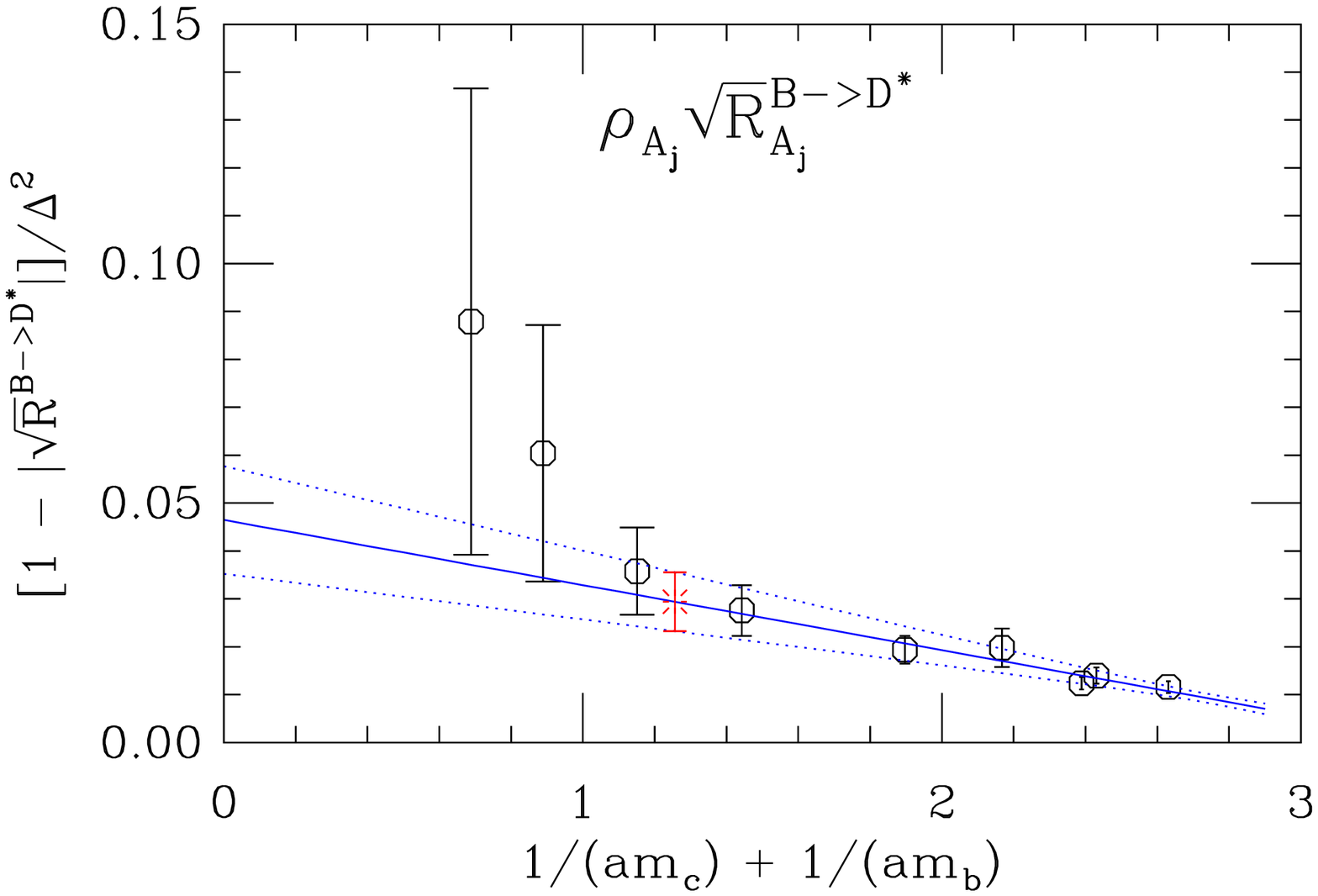,width=2.95truein}
\vspace{-8.0ex}
\caption{%
Calculated   mass dependence   for   $\sqrt{R^{B\to D^*}_{A_j}}$    at
$\beta=5.7$.  The solid line is a fit to  the expected mass dependence
given   in   \EqnRef{GenericMassDep}.   Dashed  lines show  1-$\sigma$
statistical errors from the fit.  The burst indicates the fitted ratio
at the physical combination of quark masses.}
\label{fig:ShowMdep}
\end{figure}}

\documentstyle[twoside,espcrc2,fleqn,epsfig,color,hyperref]{article}
\def\ensuremath#1{#1}
\def\myauthor[#1]#2{{#2}$^{\textrm{#1}}$}
\def\myaddress[#1]#2{$^\textrm{#1}${#2}}
\newenvironment{frontmatter}{\relax}{\maketitle}
\newdimen\mybls\mybls=\baselineskip\advance\mybls -1ex

\definecolor{Black}{cmyk}{0,0,0,1}
\definecolor{White}{cmyk}{0,0,0,0}
\definecolor{Red}{cmyk}{0,1,1,0}
\definecolor{Green}{cmyk}{1,0,1,0}
\definecolor{Blue}{cmyk}{1,1,0,0}
\definecolor{Cyan}{cmyk}{1,0,0,0}
 
\begin{document}

\begin{frontmatter}

\title{The $\BtoDstar$ Form Factor at Zero Recoil}

\author{ 
\myauthor[a]{J.N. Simone}\thanks{Presenter.},
\myauthor[b]{S.\ Hashimoto},
\myauthor[c]{A.X.\ El-Khadra},
\myauthor[a]{A.S.\ Kronfeld},
\myauthor[a]{P.B.\ Mackenzie},
\myauthor[a]{S.M.\ Ryan}
\\ [\mybls] 
\myaddress[a]{Fermilab, P.O. Box 500, Batavia, IL 60510, USA}
\\ 
\myaddress[b]{Computing Research Center, KEK, Tsukuba, 305-0801, Japan}
\\ 
\myaddress[c]{Department of Physics, University of Illinois,\
Urbana, IL 61801, USA}
} 

\begin{abstract}
We describe  a model independent  lattice  QCD method for determining
the deviation from unity for $\hAone$,  the $\BtoDstar$ form factor at
zero recoil.   We extend  the  double ratio method  previously used to
determine the $\BtoD$   form factor.   The  bulk  of statistical   and
systematic errors cancel in  the double  ratios we consider,  yielding
form factors which promise to reduce present theoretical uncertainties
in  the  determination   of $|V_{cb}|$.   We   present results  from a
prototype calculation at   a single lattice spacing corresponding   to
$\beta=5.7$.
\end{abstract}

\end{frontmatter}

\section{Introduction}

The form  factor  $\hAone$ parameterizes  hadronic  matrix elements in
$\BtoDstar$  decays.  Its  theoretical  determination is necessary  in
order to   extract the CKM     matrix  element  $|V_{cb}|$ from    the
experimental decay rate\cite{EXPT}, extrapolated to zero recoil,
\begin{eqnarray}
\lim_{\omega\to1}\frac{1}{(\omega^2-1)^{1/2}}
\frac{d\Gamma(\BtoDstar)}{d\omega}
= \qquad\qquad&& \nonumber \\
\qquad\frac{G_f^2}{4\pi^3}
(m_B-m_{D^*})^2 {m_{D^*}}^3
{|V_{cb}|}^2 |{\hAone}|^2 \FullStop &&
\end{eqnarray}


Heavy quark  symmetry constrains  this  form factor\cite{HQS}.   Up to
radiative corrections, it has deviations from unity beginning at order
$1/m_Q^2$ in   an   expansion  in   inverse  powers   of    the  quark
masses\cite{Luke}.   It is exactly one  in the infinite mass limit. We
write:
\begin{equation}
\hAone = \eta_A \left[ 1 - \delta_{1/m_Q^2} +
\mathcal{O}(\Over{\mbox{$1$}}{\mbox{$m_Q^3$}})\right] 
\;\Comma
\label{eqn:QCDhAone}
\end{equation}
where the heavy quark expansion in Heavy Quark Effective Theory (HQET)
is within brackets, and $\eta_A$  denotes radiative corrections in the
HQET-to-QCD matching\cite{Czarnecki}.

Previous calculations   of  $\delta_{1/m_Q^2}$  have relied  on  quark
models   or  sum    rules  estimates\cite{HQSreport}.   We   determine
$\delta_{1/m_Q^2}$,  in principle model  independently, using  lattice
QCD.  Our   method  extends  our  previous   work in  determining  the
zero-recoil form factor in $\BtoD$ decays\cite{BtoDpaper}.

\section{Procedure}

Consider double ratios,
\begin{eqnarray}
R^{B\to D}_{J_\mu}(t) &\equiv&
\frac{C^{D J_\mu B}(t) C^{B J_\mu D}(t)}
{C^{B J_\mu B}(t) C^{D J_\mu D}(t)} \nonumber \\
&\mathrel{\raisebox{-1.5ex}%
{$\stackrel%
{{\mathord{=}\!\mathord{=}\!\mathord{=}\!\mathord{=}\!\mathord{=}\!\Longrightarrow}}%
{{\scriptstyle T/2 \gg t \gg 0}}$}}%
&\frac{\braOket{D}{J_\mu}{B} \braOket{B}{J_\mu}{D}}
{\braOket{D}{J_\mu}{D} \braOket{B}{J_\mu}{B}}
\end{eqnarray}
of lattice three-point functions,
$C^{D J_\mu B}(t)=
\Vev{\chi_D(T/2)\;{J}_\mu ( t )\;\chi_B^{\dagger} ( 0 ) }$.
Double ratios are constructed to be identically one when the ``charm''
and ``bottom'' quarks are of equal  mass.  The bulk of statistical and
systematic uncertainties cancel in such ratios\cite{BtoDpaper}.

We need three double ratios:
\begin{eqnarray}
\RhoBar_{V_0}\sqrt{R^{B\to D}_{{V}_0}}
&\to& |h_+(1)| \,/\, \eta_{V} \label{eqn:DefPVP} \\
\RhoBar_{V_0}\sqrt{R^{B^*\to D^*}_{{V}_0}}
&\to& |h_1(1)| \,/\, \eta_{V}  \label{eqn:DefVVV} \\
\RhoBar_{A_j}\!\sqrt{R^{B\to D^*}_{{A}_j}}\;\;\;
&\!\!\!\!\!\!\!\!\!\!\!\!\!\!\!\!\!=\,\,\,\,\,\,&\!\!\!\!\!\!\!\!\!\!\!\!\!\!\!\!\!\!
\sqrt{\frac{h_{A_1}^{B D^*}\!(1)\; h_{A_1}^{D B^*}\!(1)}
{h_{A_1}^{D D^*}\!(1)\; h_{A_1}^{B B^*}\!(1)}
\frac{\eta_{A}^{DD^*}\!\eta_{A}^{BB^*}}{\eta_{A}^2}}
\FullStop
\label{eqn:DefPAV}
\end{eqnarray}
The  lattice-to-HQET matching coefficients,  $\RhoBar_{J_\mu}$,
are   near     unity  for  typical   lattice  spacings\cite{Kronfeld}.
Right-hand  expressions in the equations  above  are to be interpreted
within HQET.   Known normalizations for  elastic vector-current matrix
elements    were  used to   simplify  the   first  two  ratios.  These
normalizations are obtained nonperturbatively in our numerical work.

\TabRatioCoeffs

All three ratios have quark mass dependence
\begin{equation}
\frac{1-\RhoBar\sqrt{R_i}}{\Delta_{m_Q}^2} = c^{(2)}_i -
c^{(3)}_i\left(\frac{1}{am_c} + \frac{1}{am_b}\right)+\cdots
\label{eqn:GenericMassDep}
\end{equation}
where,         $\Delta_{m_Q}      \equiv          \left(\frac{1}{2m_c}
-\frac{1}{2m_b}\right)$.       Table~\ref{tab:MsqCoeffs}  displays our
notation for the  $c^{(2)}_i$.  Note that the  $\BtoDstar$ coefficient
contains a linear combination of the  other two coefficients. The form
of   this    coefficient is    derived   by  substituting   the   mass
dependence\cite{HQSreport},
\begin{equation}
\delta_{1/m_Q^2} =
\Delta_{m_Q}
\left(\frac{{\ell_V}}{2m_c} -
\frac{{\ell_P}}{2m_b}\right)
-\frac{{\Delta}}{4m_cm_b}
\;\Comma
\label{eqn:hA1Mdep}\end{equation}
into    the     expression for     $R^{B\to  D^*}_{{A}_j}$    shown  in
\EqnRef{DefPAV}.

Our procedure for determining ${h_{A_1}(1)}$ in $\BtoDstar$ decays is:
Extract $c^{(2)}_i$   by studying the  mass  dependence of  the double
ratios.  Solve for $\ell_P$,  $\ell_V$ and $\Delta$.  Substitute these
values and  the  values we  determine for   $m_c$ and  $m_b$ into  the
expression for  $\delta_{1/m_Q^2}$ shown in \EqnRef{hA1Mdep}.  We then
match to QCD and determine $h_{A1}(1)$ as in \EqnRef{QCDhAone}.

\section{Prototype Calculation at $\mathbf{\beta=5.7}$}

Many of the numerical details in this study are common to our study of
$\BtoD$ matrix elements\cite{BtoDpaper}. We note in particular:
{\BulletItem}We use    a subset  of   200 $\beta=5.7$  quenched  gauge
configurations on a $12^3\times 24$ lattice.
{\BulletItem}We use the  Sheikholeslami-Wohlert  quark action  with  a
tadpole-improved  tree-level coefficient,  $c_{SW}=1.57$.  Results are
interpreted within the Fermilab heavy quark formalism\cite{HeavyQ}.
{\BulletItem} We study double  ratios  for nine combinations  of heavy
quarks with bare  masses corresponding to $\kappa_h\in\{ 0.125, 0.119,
0.110, 0.100,  0.089, 0.062  \}$. Quark  masses range  around both the
charm and bottom masses.
{\BulletItem}Statistical      errors   were    obtained  using       a
single-elimination jackknife procedure.
{\BulletItem}The physical tree-level  charm and bottom  quark masses,
$m_c$ and  $m_b$, were determined  by  adjusting bare mass  inputs and
demanding that calculated meson kinetic masses  match the physical $D$
and $B$ meson masses.
{\BulletItem}Matching factors $\RhoBar_{J_\mu}$  are only known to one
loop order.  For consistency, $\eta_A$ is truncated to one loop order.
Matching factors are computed using the V scheme coupling.  We use BLM
matching scales which account for $\beta_0\alpha_s^2$ contributions.

Results in this paper  have  the spectator quark  mass fixed  near the
strange quark  mass.  In our $\BtoD$  study, we checked the dependence
upon the spectator mass for $R^{B\to  D}_{V_0}$.  Values in the chiral
limit  were consistent  with those  for the   strange quark, but  with
statistical errors  which were twice  as large.  We anticipate similar
chiral behavior for the other two ratios  we use in this study. Hence,
we  expect insignificant differences  in   $c^{(2)}_j$ in the   chiral
limit, and similar increases in  statistical errors.  Here, we account
for   the uncertainty of not   performing  the chiral extrapolation by
doubling our statistical errors.

\FigMassDep

Figure~\ref{fig:ShowMdep} shows  the  heavy  quark mass dependence  we
find for $\sqrt{R^{B\to D^*}_{A_j}}$. The quality of these results are
representative of  our  results for  the  other two ratios.    The two
points to the left in the figure have large statistical errors.  These
decays  involve the heaviest quark  masses  and suffer from well-known
signal-to-noise problems. A  fit   to the  functional  form  given  in
\EqnRef{GenericMassDep}  is  shown  in  the figure.   Its  y-intercept
shows  this   $c^{(2)}$  determination has  greater   than   $4\sigma$
significance.

We   find   this   and   the    other   two    coefficients  are    of
${\mathcal{O}}(\Lambda_{QCD})$, as expected.  The values we obtain are
broadly            consistent                 with            previous
estimates\cite{HQSreport}. Quantitative comparisons may be misleading,
however,  since uncertainties  in previous estimates  are difficult to
ascertain.

\section{Determination of $\mathbf{\hAone}$}

We present  a  \emph{preliminary} determination of $h_{A_1}(1)$  using
our prototype $\beta=5.7$ study to  illustrate the precision we expect
in a complete study:
\begin{displaymath}
h_{A_1}(1) = 0.935
\pm 0.022                 
(\,^{+0.008}_{-0.011})    
\pm 0.008                 
\pm 0.020                 
\FullStop
\end{displaymath}
Sources  of uncertainties are, respectively:
{\BulletItem}Statistical.  We double statistical errors to account for
not having extrapolated to the  down quark mass.   We must still check
the     chiral behavior of all     three  double ratios    used in the
determination of $h_{A_1}$.
{\BulletItem}Tuning of  quark  masses $m_c$  and  $m_b$.   We estimate
$10\%$ and $13\%$ uncertainties in our charm and bottom masses.
{\BulletItem}Unknown radiative  corrections beyond 1-loop. We estimate
this uncertainty by varying the 1-loop coefficients by $20\%$.
{\BulletItem}Undetermined
$\mathcal{O}(\Over{\mbox{$1$}}{\mbox{$m_Q^3$}})$   corrections      to
$h_{A_1}(1)$.      We    use   the     relative   sizes      of    the
$\Over{\mbox{$1$}}{\mbox{$m_Q^3$}}$  terms determined  in our fits  to
estimate neglected power corrections.
\\ [1.0ex]
\noindent\emph{Two important sources  of systematic  uncertainty
remain to be evaluated fully:}
{\BulletItem}Lattice spacing dependence.  In our  studies of the decay
constants  $f_D$ and  $f_B$ and of  $\BtoPi$  matrix elements we  find
lattice       artifacts  are  under  control\cite{fBpaper,BtoPi}.   We
anticipate  no large lattice artifacts  in this  study.  We adjust the
quark  action  and currents to match $\Over{\mbox{$1$}}{\mbox{$m_Q$}}$
terms of the QCD heavy quark expansion\cite{HeavyQ}.  Contributions to
$\delta_{1/m_Q^2}$  arise solely from  these terms in the double ratio
method\cite{Kronfeld}.   We note that  any remaining cutoff dependence
may  be  removed by repeating the   calculation for additional lattice
spacings and taking the continuum limit.
{\BulletItem}Uncertainty due to the   quenched approximation.  A  full
quantitative  estimate of  the error due  to  quenching must await  an
unquenched determination of $h_{A_1}$.  Note, however, the uncertainty
due to quenching affects the deviation of the  form factor from unity.
The quenching uncertainty in this deviation is commonly believed to be
perhaps $20\%$.

\end{document}